\documentclass[prl,twocolumn,showpacs]{revtex4}
\usepackage{amsmath,bbm,epsfig}
\usepackage{amsfonts}
\usepackage{mathrsfs,psfrag}
\newcommand{\ket}[1]{\ensuremath{| #1 \rangle}}
\newcommand{\bra}[1]{\ensuremath{\langle #1 |}}

\newcommand{\vac}{\ensuremath{| \mathrm{vac} \rangle}}
\newcommand{\vacbra}{\ensuremath{\langle \mathrm{vac} |}}
\newcommand{\od}[2][\alpha]{\ensuremath{\left . \hat{a} \right . _{#2,#1}}}
\newcommand{\andreaoc}[2][\alpha]{{\ensuremath{{\left . \hat{a} \right . _{#2,#1}}^{\dagger}}}}
\newcommand{\on}[2][\alpha]{\ensuremath{\left . \hat{n} \right . _{#2,#1}}}

\newcommand{\hop}[2][\alpha]{\ensuremath{\andreaoc[#1]{#2+1}\od[#1]{#2}}}
\newcommand{\snt}[2][\alpha]{\ensuremath{\on[#1]{#2} \left ( \on[#1]{#2} -1 \right )}}

\newcommand{\coefa}[1]{\ensuremath{\left . D \right._{#1}}}

\newcommand{\prob}{\ensuremath{\mathsf{P}}}
\newcommand{\hc}{\ensuremath{\mathrm{ h.c. }}}

\newcommand{\adv}{\ensuremath{V}}

\newcommand{\adwl}{\ensuremath{{U}_{1}\,}}

\newcommand{\adwx}{\ensuremath{{U}_{\times}}}
\newcommand{\jone}{\ensuremath{{J}_{1}}}
\newcommand{\jtwo}{\ensuremath{{J}_{2}}}
\newcommand{\edown}{\varepsilon_1 }
\newcommand{\eup}{\varepsilon_2 }

\newcommand{\prw}{\ensuremath{w^{\prime}}}

\begin{document} 

\title{Many-body interband tunneling as a witness for complex dynamics
in the Bose-Hubbard model}

\author{Andrea Tomadin,$^{1}$ Riccardo Mannella,$^1$ and Sandro Wimberger$^{1,2}$}  
\affiliation{$^1$Dipartimento di Fisica,
Unversit{\`a} degli Studi di Pisa, Largo Pontecorvo 3, 56127 Pisa, Italy \\
$^2$CNISM, Dipartimento di Fisica del Politecnico, C. Duca degli Abruzzi 24, 10129 Torino, Italy}

\date{\today}

\begin{abstract}
A perturbative model is studied for the tunneling of many-particle states
from the ground band to the first excited energy band, mimicking Landau-Zener
decay for ultracold, spinless atoms in quasi-one dimensional optical lattices 
subjected to a tunable tilting force. The distributions of the computed
tunneling rates provide an independent and experimentally accessible 
signature of the regular-chaotic transition in the strongly correlated 
many-body dynamics of the ground band.
\end{abstract}


\maketitle

The experimental advances in atom and quantum optics allow the experimentalist to directly study a plethora of minimal models which have been developed to describe usually much more complex phenomena occurring in solid states \cite{MO2006,BEC_mott,essl}. Bose-Einstein condensates loaded into optical lattices, which perfectly realize spatially periodic potentials, are used, e.g., to implement the Wannier-Stark problem  \cite{BEC_bloch,BEC_pisa,BEC_flo} as a paradigm of quantum transport where atoms move in a tilted lattice. Up to now all experiments on the Wannier-Stark system with ultracold atoms have been performed in a regime where atom-atom interactions are either negligible \cite{BEC_bloch} or reduce to an effective mean-field description \cite{BEC_pisa,WMMAKB2005}. State-of-the-art setups are, however, capable to achieve small filling factors of the order of one atom per lattice site \cite{BEC_mott}. Moreover, the atom-atom interactions can be tuned by the transversal confinement and by Feshbach resonances \cite{essl,BMO2003}, resulting in strong interaction-induced correlations.

\begin{figure}
  \centerline{\epsfig{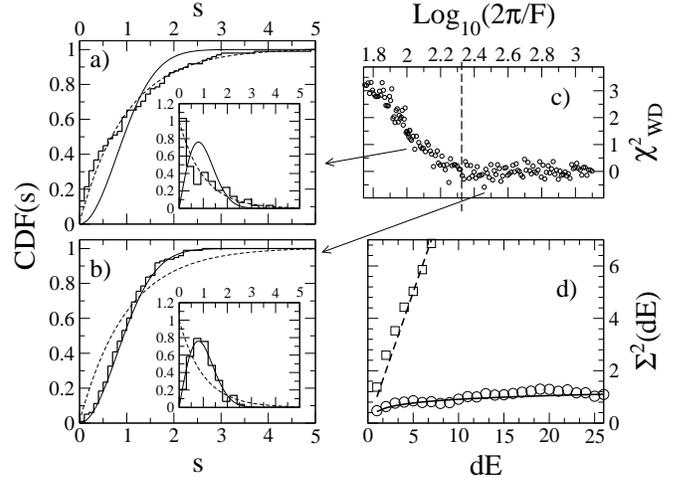}} 
  \caption{(a,b) CDF (stairs) and $\prob(s)$ (stairs in insets) 
           for $N=5$ atoms, $L=8$, lattice depth $V=10$ recoil energies
           (fixing $J_1=0.038$), $U_1=0.032$, 
           $F \simeq 0.063$ (a) and $0.021$ (b), with
           WD (solid) and Poisson distributions (dashed).
           (c) $\chi^2$ test with values close to zero for good WD statistics.
           The dashed line marks the transition to quantum chaos
           as $F$ is tuned. 
           (d) variance of the number of levels in intervals of length 
           $dE$ (with normalized mean spacing), 
           for the cases of (a) (squares) and (b) (circles), with the 
           random matrix predictions for
           Poisson (dashed) and WD (solid) \cite{mehta}.
   }
  \label{fig:1}
\end{figure}

The regime of strong correlations in the Wannier-Stark system was addressed in \cite{BK2003,KB2003}, revealing the sensitive dependence of the system's dynamics on the Stark force $F$. The single-band Bose-Hubbard model of \cite{BK2003,KB2003} is defined by the following Hamiltonian with the creation 
$\hat{a}_{l,1}^\dagger$, annihilation $\hat{a}_{l,1}$, and number operators 
$\on[1]{l}$ for the first band of a lattice $l=1\ldots L$:
\begin{equation}
\sum_{l} F l \on[1]{l} - \frac{\jone}{2} 
\left ( \hop[1]{l} + \hc  \right ) + \frac{\adwl}{2} \snt[1]{l}\;.
\label{eq:1} \end{equation}
A transition from a regular dynamical (dominated by $F$) to a quantum chaotic regime (with comparable values of $\jone, \adwl, F$) was found 
\cite{BK2003,KB2003}. The transition was quantitatively studied using the distribution of the 
spacings between next nearest eigenenergies of the Hamiltonian \eqref{eq:1}. 
This analysis \cite{BK2003,KB2003} verifies 
that the normalized level spacings $s\equiv \Delta E/ \overline{\Delta E}$ 
obey a Poisson ($\prob(s)=\exp (-s)$\,) and a Wigner-Dyson
(WD: $\prob(s)=s\pi/2\exp (-\pi s^2/4)$\,) distribution in the regular and chaotic case, respectively
\cite{mehta}. $\prob(s)$ and the cumulative distribution functions (CDF: 
$C(s) \equiv \int_0^{s}ds'\,\prob(s'))$ are shown for typical cases in Fig.~\ref{fig:1}, where we scanned $F$ to emphasize the crossover between the regular and the chaotic regime. Statistical tests are also
shown which confirm the analysis of \cite{BK2003,KB2003} in a more systematical manner 
\cite{andrea}.

As shown in \cite{BK2003}, the strong correlations in the quantum chaotic regime induce a fast and 
irreversible decay of the Bloch oscillations, which otherwise would persist in the ideal, non-interacting
case. Therefore, the crossover between the two regimes discussed above could be measured in 
experiments by observing just the mean momentum as a function of time. Here we introduce a new, robust
and hence also experimentally accessible prediction for this crossover. 
In the presence of strong interactions 
parameterized by $\adwl$, the single-band model should be extended to allow for interband transitions
\cite{SS2005}, as recently realized at $F=0$ in experiments with fermionic interacting atoms \cite{essl}.
Instead of using a numerically hardly tractable complete many-bands model, we introduce a perturbative
decay of the many-particles modes in the ground band to a second energy band. Our novel approach to study the Landau-Zener-like tunneling between the first and the second band \cite{MO2006,BEC_pisa,WMMAKB2005,GKK2002,mean} leads to predictions for the expected decay rates and their statistical distributions. As we will show, the latter are drastically affected by the dynamics in the ground band, and they therefore provide a measurable witness for the regular-chaotic transition.

We first derive the individual decay rates
of the dominating interband coupling channels. These decay rates will serve to effectively open the 
single-band model (\ref{eq:1}) for mimicking losses arising from the interband coupling.
Our analysis starts from the following ``unperturbed'' Hamiltonian for the first two bands:
\begin{eqnarray}
H_{0} & = \sum_{l=1}^{L} \left[ \edown \on[1]{l} + \eup \on[2]{l} 
        - \frac{\jtwo}{2} ( \hop[2]{l} + \hc ) \right. \nonumber \\ 
& \left. + F l \left( \on[1]{l}+\on[2]{l}  \right) + \frac{\adwl}{2} \snt[1]{l} \right] \;.
\label{hamupt}
\end{eqnarray}
For a moment, we neglect the hopping in the lower band, where the single-particle Wannier functions
\cite{GKK2002} are more localized than in the upper band. In the latter we neglect the interactions, 
since initially only a few particles 
populate the excited levels. A closer analysis of the full two-bands system \cite{andrea} 
shows that there are {\it two} dominating mechanisms that promote particles to the second band.
The first one is a single-particle dipole coupling arising from the force term:
\begin{equation}
H_{1} = F \cdot \coefa{} \sum_{l}\left ( \andreaoc[2]{l}\od[1]{l} + \andreaoc[1]{l}\od[2]{l}  \right ) \;,
\label{perone}
\end{equation}
where $\coefa{}$ depends only on the lattice depth $V$ (measured in recoil energies
according to the definition in \cite{WMMAKB2005}). The second one is a many-body effect, 
describing two particles of the first band entering the second band together:
\begin{equation}
H_{2} =\frac{\adwx}{2} \sum_{l=1}^{L}\left (
 \andreaoc[2]{l}\andreaoc[2]{l}\od[1]{l}\od[1]{l} + (1 \leftrightarrow 2)  \right ).
\label{pertwo}
\end{equation}
The cross-band interaction is characterized by the parameter
$\adwx \equiv {\tilde a}_s \int dx\,\chi_1^2\chi_2^2 \simeq 0.5 \adwl$ 
(for $V = 3\ldots 10$) \cite{andrea}, for $\adwl \equiv {\tilde a}_s \int dx\,\chi_1^4$, with
renormalized scattering length ${\tilde a}_s$ \cite{BMO2003,andrea} and the Wannier functions $\chi_{1,2}$ localized
in each well for the first or second band. To justify the following perturbative approach, it is crucial to realize 
that the terms \eqref{perone} and (\ref{pertwo}) must be small compared with the band gap 
$\Delta \equiv \eup - \edown$ (not necessarily small with respect to the single band terms in \eqref{eq:1}), 
and indeed $FD, \adwx, \adwl \ll \Delta$ for the parameters considered here.

For the first perturbation, the decay channel of a given unperturbed Fock state labelled
$\ket{b}$ (with a total number of atoms $N$ and $n_h$ atoms in an arbitrary well $h$) is
\begin{equation}
\ket{b;N}\otimes \vac\rightarrow \ket{b';N-1}\otimes \ket{w}\;,\;n'_{h}=n_{h}-1.
\label{dcnone}
\end{equation}
Here, $\ket{w} = \sum_{m=-\infty}^{+\infty} \mathcal{J}_{m-w} (|\jtwo | / F ) \andreaoc[2]{m} \vac $ is 
the single-particle eigenstate for the Wannier-Stark problem, localized around the site $w$ in the second band, 
with the Bessel function of the first kind $\mathcal{J}_{m}(x)$ \cite{GKK2002}.

The expectation value of \eqref{perone} for $\ket{b;N}$ of the first band, equal to the first-order $\delta E(b)$, is zero because the operator does not conserve the number of particles within the bands.
The decay width at first-order is given by the matrix element of the perturbation between the initial and final state according to Fermi's Golden Rule, and only the first term in \eqref{perone} gives a 
nonzero contribution \cite{andrea}:
\begin{eqnarray}
&\bra{k}\bra{b'}\sum_{l=1}^{L} \andreaoc[2]{l}\od[1]{l} \ket{b}\vac = \sum_{l=1}^{L}\mathcal{J}_{l-w}( |\jtwo |/F)
\nonumber \\
&\cdot \delta(n'_{l},n_{l}-1)\;\sqrt{n_{l}} \;\prod_{m\neq l} \delta(n'_{m},n_{m}).
\end{eqnarray}
The $\delta(\cdot,\cdot)$ functions act as a selection rule for the Fock states that are coupled by the perturbation. The tunneling mechanism does not include any income of energy from an external source, so the initial and final energies
$E_{0}(b)  =  \vacbra\bra{b}H_{0}\ket{b}\vac$ and
$E_{0}(b',w)  =  \bra{w}\bra{b'}H_{0}\ket{b'}\ket{w}$, respectively,
must be equal as required by the Golden Rule.
The condition on the energy conservation is, however, relaxed to account for the uncertainty $\Delta E(b)$ of the unperturbed energy levels of the initial and final states in the lower band arising from the
hopping in this band initially neglected in \eqref{hamupt}.
A detailed derivation is given in \cite{andrea}, and here we only state the result:
\begin{eqnarray}
\Delta E(b) = 2\pi \left ( \jone \,/2 \right )^{2}\; \sum_{b'}\,\Delta E (b\rightarrow b') =
\nonumber \\ 2\pi \left ( \jone \,/2 \right )^{2}\; \sum_{l}\,\sum_{\Delta l=\pm 1}
n_{l}^{2}\; \delta(n_{l+\Delta l}+1,n_{l}).
\end{eqnarray}
The level density $\rho(E,b)$ around the unperturbed energy $E_{0}(b)$ of a Fock state $\ket{b}$ 
is then approximated by a rectangular profile, of width $\Delta E(b)$ and unit area:
$\rho(E,b) =  \chi \left \lbrace |E-E_{0}(b)|\leq \Delta E(b) / 2 \right \rbrace\; /\; \Delta E(b)$.
The relaxed energy conservation rule selects from \eqref{dcnone} the set $K$ of 
permitted decay channels $(h,w)$ parameterized by the two indices $h,w$ such that:
\begin{eqnarray}
& E_{0}(b',w)-E_{0}(b) = \Delta -F (h-w)-\adwl\left ( n_{h}-1 \right ) \nonumber \\
& \in \left [-\frac{\Delta E(b)+\Delta E(b')}{2},\,\frac{\Delta E(b)+\Delta E(b')}{2} \right ].
\end{eqnarray}
Hence the energy $\Delta$ required to promote a particle to the second band  is supplied by the decrease of the interaction ($\propto \adwl$) and by the work of the force ($\propto F$) exerted on the promoted particle.

The total width $\Gamma_{1}(b)$ for the decay via the allowed channels $K$, 
is proportional to the square of the matrix element and to the level density $\rho(E,b)$:
\begin{eqnarray}
& \Gamma_{1}(b)  =  2\pi (FD)^2 \sum_{(h,w) \in K} \left \lbrace
\left | \mathcal{J}_{h-w} (\frac{|\jtwo |}{F}) \cdot \sqrt{n_{h}} \right |^{2} \cdot 
\right . \nonumber \\ & \left . \frac{1}{\Delta E(b)\Delta E(b')} \right \rbrace.
\label{gampea}
\end{eqnarray}
$\mathcal{J}_{m}(x)$ significantly contributes
only for $|m|\lesssim |x|$.
If $\adwl, \Delta E(b) \ll \Delta$, the energy conservation is roughly given by $|\Delta| \simeq F (h-w)$. 
Requiring that the Bessel function in \eqref{gampea} is substantially larger than zero, we obtain the inequality $|\Delta| \leq |\jtwo |$. The last condition does not depend on $F$, since a twofold effect is at work: a stronger force produces a larger energy gain when a particle moves along 
the lattice, but the extension $|\jtwo /F|$ of the single-particle state shrinks.
Therefore, increasing $F$ results in an increased energy matching and a strongly reduced ``geometrical'' matching. For $3 < \adv < 26$, we have $|\Delta|-|\jtwo |>1.0$ \cite{andrea}, such that the energy matching cannot be realized by just tuning the lattice depth. The decay can, however, be activated by an increase of the interactions, which can be experimentally achieved by acting on the transversal confining potential of a quasi-one dimensional lattice, or by a Feshbach resonance \cite{BMO2003}. 
In the calculations presented below, we augmented $\adwl$ used in \cite{BK2003,KB2003} 
by a factor of order $10$, and as noted in the introduction, a similar increase of the 
interaction strength was used in the experiment
to promote fermions to higher bands \cite{essl}, in close analogy to the here 
described field- {\it and} interaction-induced interband coupling of bosons.

The second term \eqref{pertwo} is treated in a similar way, with the difference that two particles are promoted to the second band, and the position of the second single-particle state $\ket{\prw}$ 
is an additional degree of freedom for the transition. The decay channels are:
\begin{equation}
\ket{b,N}\otimes\vac\rightarrow \ket{b',N-2}\otimes\ket{w,\prw} \,;\, n'_{h}=n_{h}-2.
\end{equation}
The energy matching selects a set $K$ of decay channels, 
parameterized by the three site indices $h,w,\prw$:
\begin{eqnarray}
(h,w,\prw)\;\in \;K\;\mbox{ s.t. }\quad E_{0}(b',w,\prw)-E_{0}(b)= \mbox{} \nonumber \\
\mbox{} = 2 \Delta - F (2h-w-\prw)-\adwl\left ( 2n_{h}-3 \right ) \nonumber \\
\in\; \left [-\frac{\Delta E(b)+\Delta E(b')}{2},\,\frac{\Delta E(b)+\Delta E(b')}{2}   \right ].
\end{eqnarray}
The computation of the matrix element yields \cite{andrea}: 
\begin{eqnarray}
& \Gamma_{2}(b) =  2\pi \left ( \frac{\adwx}{2} \right )^{2} 
\sum_{(h,w,\prw)\in K} \left \lbrace 
\left | \mathcal{J}_{h-w}(\frac{|\jtwo |}{F}) \; \cdot\right . 
\right . \nonumber \\ & \left . \left . \mathcal{J}_{h-\prw}(\frac{|\jtwo |}{F}) \right |^{2} \,\cdot \, 4 n_{h} \left (  n_{h}-1   \right ) \frac{1}{\Delta E(b)\Delta E(b')} \right \rbrace.
\label{gampeb}
\end{eqnarray}
With respect to \eqref{gampea}, the additional degree of freedom $\prw$ 
results in a summation over all possible values of $w-\prw$. 
This follows from the possibility to conserve the energy even if a particle is pushed far, if the other particle is pushed almost equally far in the opposite direction.
Since the decay widths in \eqref{gampeb} depend on the product of two (rapidly decaying) Bessel functions -- again a ``geometrical'' matching condition -- we apply the truncation $|w-\prw| \leq |\jtwo /F|$, to reduce the formula to a finite form.

\begin{figure} 
  \centering 
  \includegraphics[width=\linewidth]{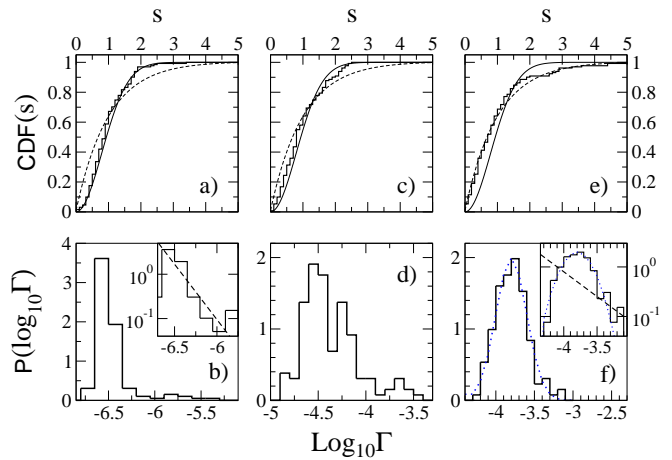}
  \caption{
(a,c,e) CDF from $\mathsf{Re}\,\{E_{j}\}$ (stairs), 
together with WD (solid) and Poisson predictions (dashed). 
(b,d,f) distributions of the logarithm of the rates. 
In (a,b), (c,d), (e,f) $F \simeq 0.17,0.31,0.47$, respectively, with
$(N,L)=(7,6), V=3, \adwl = 0.2$ (fixing $\adwx \simeq 0.1)$.
In the regular regime (f), a log-normal distribution (dotted) 
well fits the data, with a scaling 
P$(\Gamma) \propto \Gamma^{-x}$ for the largest
$\Gamma$ (dashed line in the inset of (f) with $x=1$). In the chaotic case, a global 
power-law behavior with $x \approx 2$ is found (dashed line in the inset of (b)).
}
  \label{fig:gamsca} 
\end{figure} 

\begin{figure} 
  \centering 
  \includegraphics[width=\linewidth]{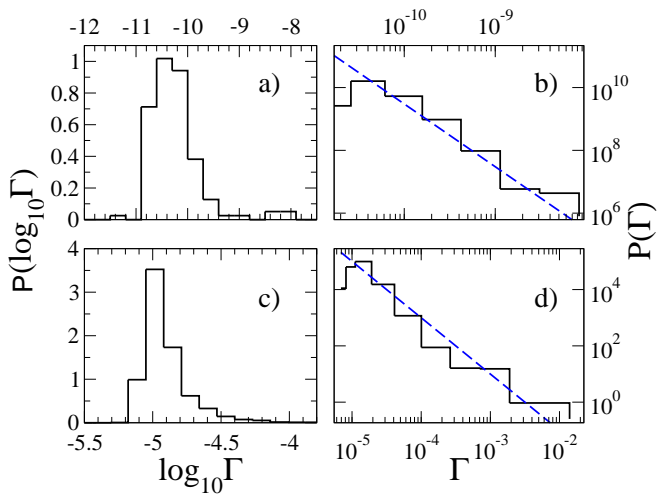}
  \caption{(a,c) rate distributions in the chaotic regime with
           $F \simeq 0.17$, $\adwl = 0.2$ ($\adwx \simeq 0.1$), together with
           the corresponding {\em unscaled} P$(\Gamma)$ in (b,d).
           In (a,b) $(N,L)=(7,6), V=4$, and in (c,d) $(N,L)=(9,8), V=3$.
           Power-laws P$(\Gamma) \propto \Gamma^{-x}$ are found with $x \approx 2$ (dashed lines
           in (b,d)).
}
  \label{fig:gamscb} 
\end{figure}

We can now compute the total width $\Gamma_{\mathrm{F}}(b)=\Gamma_{1}(b)+\Gamma_{2}(b)$ defined by the
two analyzed coupling processes for each basis state $\ket{b}$ of the single-band problem given in 
\eqref{eq:1}. The $\Gamma_{\mathrm{F}}(b)$ are inserted as complex potentials
in the diagonal of the single-band Hamiltonian matrix.
After a gauge transform that recovers the translational invariance of the problem (see \cite{KB2003,andrea} for details), the latter matrix is used to compute the evolution operator over one Bloch period 
$T_{\mathrm{B}}$,
which is finally diagonalized to obtain its eigenphases 
$\exp\left( -iE_{j}\,T_{\mathrm{B}} \right )$. Along with the statistics of the level spacings
defined by $\mathsf{Re}\,\{E_{j}\}$, the Figs. 
\ref{fig:gamsca} and \ref{fig:gamscb} analyze the statistical distributions of the tunneling rates
$\Gamma_{j}=-2\mathsf{Im}\,\{E_{j}\}$ for some paradigmatic cases.
All rates are much smaller than unity, which a posteriori is fully 
consistent with our perturbative approach.

To observe what happens at the regular-chaotic transition (c.f. Fig.~\ref{fig:1}), 
we scan $F$ in Fig.~\ref{fig:gamsca}, and as $F$ increases, the average decay 
increases by orders of magnitude, while the distributions broaden. 
The large increase of the rates is due to an improved energy matching, when $F$
supplies the necessary energy to promote particles to the second band. For the parameters 
of Fig.~\ref{fig:gamsca}, the single-particle Landau-Zener formula \cite{GKK2002} gives 
$\Gamma_{\mathrm{LZ}} = F/(2\pi) \exp \left [ -\pi^2 \Delta^{2} /(8 F)  \right ]
\sim 10^{-23},\, 10^{-12},\, 10^{-8}$ for (b,d,f).
This huge variation, typical of semiclassical formulae, 
implies that there are possibly parameters for which our results are comparable to 
the single-particle prediction, but, in general, the many-particle effects {\it cannot} be neglected.
Moreover, mean-field treatments of the Landau-Zener tunneling at best predict a shift of $\Gamma$
\cite{WMMAKB2005,mean}, but cannot account for their distributions.

In the chaotic regime, the Fock states are strongly mixed by the dynamics \cite{BK2003,KB2003,andrea} 
and a fast decaying Fock state can act as a privileged decay channel for {\em many} 
eigenstates. Many states then share similar rates, leading to thinner
distributions. Therefore, the thinner distribution of Fig.~\ref{fig:gamsca} (b) 
is a direct signature of the chaotic dynamics evidenced in (a), as compared with 
the regular case in (e,f). In Fig.~\ref{fig:gamsca} (f), we found a good agreement
with the expected log-normal distribution of decay rates \cite{TG2000} (or of the
similarly behaving conductance \cite{been97}) in the regular regime. 
There the system shows nearly perfect Bloch
oscillations \cite{BK2003}, and the motion of the atoms is localized along the 
lattice \cite{GKK2002}. We can even detect a qualitative 
crossover to a power-law P$(\Gamma) \propto \Gamma^{-1}$
in the right tail of the distribution, as predicted from localization theory \cite{TG2000,gamma,Kottos}.
The distributions in Figs.~\ref{fig:gamsca} (b) and \ref{fig:gamscb} follow the
expected power-law for open quantum chaotic systems in the diffusive regime 
\cite{Kottos}. The exponents $x$ are, however, nonuniversal and depend on the opening of 
the system. In our case, the decay channels are defined by the interband coupling, which in a sense 
attaches ``leads'' to {\em all} lattice sites {\em within} the sample. 
Going along with the regular-to-chaotic transition in the lower band
of our model (from Fig.~\ref{fig:gamsca} (f) to (b), or to Fig.~\ref{fig:gamscb}) the $\Gamma$
distributions transform from a log-normal to a power-law with $x \approx 2$, in close analogy to the 
transition from Anderson-localized to diffusive dynamics in open disordered systems \cite{Kottos,jpa}.

In summary, our perturbative opening of the single-band
Wannier-Stark system allows one to study Landau-Zener-like interband tunneling within a many-body description of the dynamics of ultracold atoms. The 
statistical characterization of the tunneling rates
(mean values and form of the distributions) provides clear and robust 
signatures of the regular-to-chaotic transition for future experiments. 
A more detailed analysis of the interband coupling in a full-blown model, in which at least two 
bands are completely included, calls for huge computational resources to access the complete quantum 
spectra. Nonetheless, our results are a first step in the direction of studies for
which ``horizontal'' and ``vertical'' quantum transport along the lattice are 
simultaneously present and influence each other in a complex manner.

We thank the Centro di Calcolo, Dipartimento di Fisica, 
Universit\`a di Pisa, for providing CPU, and the Humboldt Foundation, 
MIUR-PRIN, and EU-OLAQUI for support.

\end{document}